\RequirePackage{fix-cm}
\documentclass[twocolumn]{svjour3}          % twocolumn

\smartqed  % flush right qed marks, e.g. at end of proof
\usepackage[final]{pdfpages}
\usepackage{graphicx}
\usepackage[numbers]{natbib}
\usepackage{comment}

\usepackage{draftwatermark} 
\SetWatermarkFontSize{1cm}
\SetWatermarkScale{5}
\SetWatermarkText{\sf PREPRINT}
\SetWatermarkLightness{.80}
\def\makeheadbox{\relax}
\begin{document}
\newcommand{\textcite}[1]{\citeauthor{#1} \cite{#1}}

\title{Confident yet Concerned: Inconsistencies in Computing Students' Attitudes
on Cybersecurity %\thanks{Grants or other notes
}


\author{Victor Adama \and Robert Biddle \and Nalin Arachchilage \and Danielle Lottridge}


\institute{V. Adama \at
              University of Auckland
              \email{avic888@aucklanduni.ac.nz}           %  \\
           \and
        R. Biddle \at Carleton University
              \email{robert.biddle@carleton.ca}
            \and
           N. Arachchilage \at The Royal Melbourne Institute of Technology
              \email{nalin.arachchilage@rmit.edu.au}
            \and
           D. Lottridge \at University of Auckland
              \email{d.lottridge@auckland.ac.nz}
}

\date{Received: August 2025 / Accepted: May 2026 --- International Journal of Information Security}

\maketitle

\begin{abstract}
Today's young adults are most immersed in technology, leading in feelings of powerlessness in managing online \textcolor{black}{privacy across many platforms, and particularly susceptible to phishing attacks. This raises questions about their general, wide-ranging attitudes towards and management of cybersecurity.} How do young, tech-savvy adults approach cybersecurity? We seek a better understanding of their \textcolor{black}{cybersecurity knowledge, attitudes and experiences, in particular in} addressing deceptive online communications. We surveyed a group of `lead users': computing university students (\textit{n} = 236). By combining thematic analysis of open-ended responses with quantitative data, we provide insights into their experiences and perceptions. While students demonstrate reasonable cybersecurity awareness, their cybersecurity experiences vary, and inconsistencies exist around their practices, perceptions of responsibility, and support structures. Findings also reveal four key thematic tensions: 1) Computing students are knowledgeable yet have persistent incorrect beliefs, 2) They learn more about keeping safe from sources outside the classroom, 3) They have limited assistance and have fallen victim to cybercrime, and 4) Many are confident, yet others are concerned about their own safety and responsibility. Through cluster analysis of attitudes, we identify two groups, with \textcolor{black}{one feeling less prepared, less confident, yet expressing a desire to learn more. Established measures of intentions and objective knowledge were correlated to preparedness. Self-efficacy correlated to confidence and predicted cluster membership.}

%\textbf{Design/Methodology/Approach} - 
%\textbf{Findings} - 

%\textbf{Purpose} -

\end{abstract}

%%%%%%%%%%
%\section{Introduction}
%\label{intro}
%Your text comes here. Separate text sections with
%\section{Section title}
%\label{sec:1}

%\subsection{Subsection title}
%\label{sec:2}
%as required. Don't forget to give each section
%and subsection a unique label (see Sect.~\ref{sec:1}).
%\paragraph{Paragraph headings} Use paragraph headings as needed.
%%%%%%%%%%%

%%%% ---- PAPER STARTS HERE

\section{Introduction}
\label{sec:introduction} 
Cybersecurity is hard \cite{kovavcevic2020sawit}. However, we would expect tech-savvy young adults in tertiary education, particularly students studying computing, to be some of the best equipped to handle it \cite{venter2019cyber}. Yet, the technical aspects they understand---such as antivirus software, firewalls, and spam filters---have over time proven insufficient to address all cybercrime \cite{furnell2009recognising,parsons2014determining, jeong2019towards}, leaving them vulnerable \cite{aliyu2010computer,fredericks2016comparing,victor2018password, filippidis2018information,moallem2019cybersecurity}.  In our study, we focus on computing students as `lead users' \cite{von1986lead} to understand contemporary reactions and attitudes to today's cyber climate, and in particular toward social engineering such as phishing, which cuts past technical defences to be today's most effective and pervasive cyber tactic \cite{abawajy2014user,aldawood2018educating,wang2018framework}.

Young people, especially those studying computing, are deeply embedded in digital ecosystems where their personal data is constantly collected and monitored. 

The framing of our study is what Draper and Turow \cite{draper2019corporate} refer to as ``digital resignation''.  They suggest that is a rational response of some people to the widespread monitoring and collection of personal data online: ``they are resigned'' ---``they are convinced that surveillance is inescapable''. In 2006, Barnes \cite{barnes2006privacy} used the term ``privacy paradox'' to describe how young people despite disagreeing that ``everybody should know everything about everyone else'', they nevertheless used online platforms for social connections, seemingly unaware that the platforms were public spaces. Much research has followed on general relationships between privacy intent and actual disclosure, and is reviewed and discussed by Kokolakis \cite{kokolakis2017privacy}. Like Barnes, Draper and Turow focus on online platforms, and they suggest people are not necessarily uninformed, nor making careful rational trade-offs. Rather, the reason for the apparent contradiction is that people perceive the power of online platform providers as ``inevitable and immovable feature of contemporary life''. Their proposal is for a theoretical framework for understanding the phenomenon, based on earlier work on the concept of resignation, addressing feelings of helplessness in crisis, and of futility and cynicism in the face of powerful external influences.  Hoffmann et al. described the experience as privacy cynicism, defined as ``an attitude of uncertainty, powerlessness and mistrust towards the handling of personal data by online services, rendering privacy protection behaviour subjectively futile'' (p. 2) \cite{hoffmann2016privacy}.

Those theories all concern \textit{privacy}, meaning access by online platforms to personal information. It is possible that the complex factors shaping paradoxical attitudes and behaviour around privacy may also be affecting security \cite{stanton2016security}, since young adults are seeing as leading as well as vulnerable. Given the difficulty and ever-evolving nature of cybersecurity, the close overlap between privacy and security, this motivates a broad-ranging examination of tech savvy young people's security knowledge (what they know about phishing-related attempts), the practical steps they take to address phishing-related attempts (know-how), their attitudes towards today's cyber climate, and their experiences in addressing phishing. 
\textcolor{black}{
This is an important issue, because if young people studying computing are resigned to cybersecurity vulnerabilities, then we should reconsider cybersecurity support and education.}
We ask the following research question:

\begin{itemize}
    \item What are computing students’ abilities, attitudes, and experiences around deceptive online interactions, particularly toward phishing?
\end{itemize}

The complex interplay we \textcolor{black}{anticipate with security} (as observed with privacy) led us to an exploratory design approach. While a strong theoretical framing and engagement with frameworks are often ideal, an overemphasis on pre-existing theory can sometimes hinder the discovery of critical insights, especially when the phenomenon under study is still emerging, dynamic, or inadequately theorized. In considering how the field of Information Systems (IS) tends to link empirical data collection to theory, Fink emphasises the need to “leave some room for exceptions\dots that may yield long-term benefits”  (p. 98) \cite{fink2021philosopher} when researchers encounter “interesting and consistent evidence that cannot be deductively derived from established theory” (p. 98) \cite{fink2021philosopher}. In line with this reasoning, our study adopted a discovery-oriented approach. Rather than forcing the data into a predetermined theoretical framework, we allowed the insights to surface inductively, so that future research efforts could draw stronger theoretical and practical connections. In particular, we attempted to be non-judgemental and avoid emphasis on ``compliance''.  

To assess a reasonably sized sample of computing students, we employ a survey approach.  We employ factor analysis to show the strongest concepts, cluster analysis to identify distinct groups of participants\textcolor{black}{, and logistic regression for inferential analysis.}  % May be update here with recent analysis
The results show a mixed picture, with fair knowledge and some confidence, but simultaneously a range of inconsistencies and concerns. We observed that those with more enthusiasm felt less prepared and confident, and those with less enthusiasm felt more prepared and confident. We discuss computing students' emerging contemporary attitudes towards cybersecurity as well as implications for cybersecurity education. Our study contributes: 
\begin{itemize}
    \item empirical findings on computing students’ cybersecurity abilities, attitudes, behaviours, and experiences, and inductively identified themes that reveal inconsistencies across these dimensions.
    \item \textcolor{black}{identified subgroups that differ in preparedness, confidence and interest. Cluster membership is associated to established scales for cybersecurity intentions (SeBIS), objective knowledge (PEW), word-of-mouth sources, and cluster membership is predicted by self-efficacy.}

\end{itemize}

\section{Related Work}
%\label{sec:Related Work}
\label{sec:relatedwork}

Research on computing students' cybersecurity abilities reveals \textcolor{black}{gaps} between theoretical knowledge and actual practice. A 2024 survey of 126 undergraduate students in computing-related programs at the University of Colorado, Boulder found that they acknowledged the importance of cybersecurity but lacked understanding of what it entails. They perceived confidentiality, integrity, and availability as more relevant than cybersecurity itself and thought that studying cybersecurity requires advanced maths skills \cite{cowit2024computing}. A 2018 study of 26 computing students at a range of U.S. universities using think-aloud interviews revealed confusion between  cybersecurity concepts, such as authentication and authorization, and also about encryption and the role of digital certificates. Notably, these misconceptions persisted even among students who had completed cybersecurity courses \textcite{thompson2018student}. A 2019 survey of 247 students from two universities in Silicon Valley found only 26\% agreed or strongly agreed that they were knowledgeable about cybersecurity. The study revealed a troubling contradiction: students acknowledge the risks associated with digital communications but lack adequate strategies for safeguarding their information. Despite being in a tech-savvy region, some participants displayed insufficient awareness of data protection measures such as two-factor authentication (2FA) and the risks of online surveillance. About 8\% did not know what two-factor authentication was, and only a minority demonstrated secure behaviours such as using two-factor authentication consistently or employing strong passwords across accounts.  Moallem argued that educational institutions fail to adopt proactive strategies to equip students with the necessary skills to mitigate cybersecurity  threats\cite{moallem2019cybersecurity}, echoing findings from \textcite{cowit2024computing}. 

Studies from other countries show similar results. A survey of 87 students at a Greek university examined how higher educational levels correlated with greater cybersecurity awareness and adherence to good cybersecurity practices and ethical standards \cite{filippidis2018information}. Masters level students demonstrated greater awareness and compliance compared to their undergraduate counterparts. Despite this, there were noticeable gaps in awareness levels. Students specialising in computing exhibited a higher degree of technical knowledge, but also demonstrated a tendency to  engage in insecure practices. Their adoption of tools and protective measures was noted to have been minimal and more than half used the same password across multiple accounts. 

Other research efforts examined how computing-related students manage passwords. In a 2016 survey of 45 South African university computing students, while participants demonstrated awareness of theoretical knowledge on password policies, many did not implement these principles consistently \cite{fredericks2016comparing}. Many students reused passwords across multiple accounts, failed to delete unused accounts, wrote passwords down and engaged in other insecure practices. A 2018 study with 481 computing students at a Nigerian university echoed these findings: there was widespread awareness of good password practices but notable failures to apply them in everyday practice \cite{victor2018password}. Both studies emphasised that awareness alone was insufficient to ensure protection. 

The issue of responsibility emerged as a significant theme in a 2015 study of 274 students enrolled in an introductory Information Systems course. Findings revealed that many students viewed cybersecurity as a technical responsibility associated with IT professionals, rather than something that required their active involvement \cite{pawlowski2015social}. Their social representations of cybersecurity tended to focus on external threats like hackers, while personal actions and behavioural risks were often overlooked. The authors argue for instructional design that promotes personal responsibility and contextual understanding, rather than focusing solely on abstract or technical aspects.

Research efforts have also compared computing students to peers from other disciplines. A 2023 study comparing cybersecurity practices between computing and other students at a large U.S. university \cite{cravens2023comparison} found that while both groups demonstrated similar levels of practical cybersecurity knowledge, computing students exhibited significantly better password hygiene. Key differences included stronger use of random password elements, more frequent password updates, and higher adoption of two-factor authentication among computing students.  Non-computing students were more likely to engage in insecure practices such as using identical or predictable passwords \cite{cravens2023comparison}.  In contrast, a 2023 study with 1,710 students from eight Chinese universities (grouped into computing, science and engineering, and liberal arts majors) expected that those with computing majors would outperform their peers, however results revealed that all students showed weak password practices regardless of their major. The authors reported that differences in cybersecurity awareness across both categories were largely insignificant\cite{guo2023survey}. 

In summary, while computing students often display marginally better awareness of cybersecurity threats and ethical issues than their non-computing peers, this advantage is not consistent or comprehensive. Across studies, a recurring pattern of knowledge–practice gaps and limited engagement with preventive behaviour is observed. While more general research efforts reveal that young adults are particularly susceptible to phishing attacks \cite{sheng2010falls,klutsch2024friend,algarni2017empirical}, computing students—a subset of this demographic—are especially compelling to examine. This is mainly because despite their formal exposure to information technology concepts and assumed familiarity with digital environments, the potential paradox between their technical inclination and actual cybersecurity behaviour warrants further attention. Their relative advantage in awareness over non-computing peers is neither consistent nor sufficient to improve safety, underscoring the need to understand the complexities  shaping their cybersecurity posture.

\section{Method}
\label{sec:Method}

To better understand computing students' abilities, attitudes, and experiences around cybersecurity and deceptive online interactions, we employed a mixed-methods survey \cite{ivankova2006using}.  

\subsection{Participants}

We recruited 236 participants from a non-cybersecurity computing course at the University of Auckland, an urban university in an OECD country with a Computer Science department that is ranked within the top 100 worldwide.  Participants were recruited from undergraduate Human-Computer Interaction (HCI) courses comprising Computer Science (n = 255) and Software Engineering (n = 86) students. Participation was voluntary, and approximately 69\% of students chose to take part. As the sample was drawn from an HCI course, it is likely participation was skewed towards students with interests in HCI. However, they were all final year students, and all would have passed earlier courses in programming, algorithms, computer organisation, and mathematics, and a range of more specialised computing courses. Participants were offered course credit for participating, along with the opportunity to win a \$200  voucher. 68\% of participants were male, 26\% female, and 6\% preferring not to say. The age distribution was between 16--24 (93\%), 25--34 (6\%) and 35--44 years old (1\%). Approximately 5\% had full-time jobs, 43\% had part-time jobs, 3\% were self-employed, and 37\% were not employed.  12\% described their employment,  reflecting jobs across various sectors, including sales, events, training, finance, IT, and others. Most owned and used various devices (phones 98\%, computers 99\%, tablets 49\%) and accessed the Internet daily (99.5\%).

\subsection{Survey Design}
\label{surveydesign}
The survey design followed a process of question development, expert validation with five cybersecurity experts, and pilot testing. A pilot study with a sample of 20 participants was conducted (not included in the final tally of 236). The results highlighted areas where wording was unclear, and response options were limited, enabling us to refine the survey for clarity, relevance, and comprehensiveness. 
\textcolor{black}{
\subsection{Ethical Considerations}
Our University requires a comprehensive explanation of ethical considerations be provided before granting clearance for research studies such as ours. We first explained the importance of online fraud and deception such as phishing, and that research has shown the importance of education and training to avoid the danger, referencing Rahman et al. \cite{rahman} and Gupta et al. \cite{gupta}. We provided a participant information sheet which explained the purpose of the study, assuring that participation was voluntary and would not affect relationships with the University. Data management procedures to ensure confidentiality were also explained. We offered contact information for support, especially offering cultural support for Māori (the indigenous population of New Zealand). The study received approval from the university's human research ethics committee. We address limitations in this process at the end of this paper.
}
\subsection{Survey Structure}
We outline below the structure and flow of the survey. See the appendix for the full survey questions. The survey had an estimated completion time of 28 minutes, as assessed by the Qualtrics survey platform.

\paragraph{Initial Information, Consent and Inclusion Criteria}
The survey began with a brief introduction to the intent of the study. A link to a Participant Information Sheet was provided. A consent form described rights and options regarding participation. 

\paragraph{Cybersecurity Knowledge and Experiences}
Questions included self-reported IT knowledge, understanding of cybersecurity, perceived ability to recognise deceptive communications (emails, text messages and phone calls), and experiences with cybercrime. 

\paragraph{Sources of Cybersecurity Knowledge}
Open-ended questions asked about the sources of cybersecurity knowledge.
This section asked questions to establish knowledge acquisition sources, either through formal education, self-directed learning, online resources, personal experiences, and informal support networks. 

\paragraph{Attitudes Toward Cybersecurity}
Questions in this section \textcolor{black}{asked about their attitudes relating to cybersecurity, its importance }and their level of concern about cyber risks. \textcolor{black}{We also asked them to assess their confidence and ability in managing cybersecurity threats.} These questions aimed to capture the participants' mindset regarding online safety and their perceived responsibility in protecting themselves. 

\paragraph{Technology Usage}
To contextualise participants' cybersecurity practices, the survey included questions about technology usage: frequency of use, types of devices and platforms, and online activities.

\paragraph{Cybersecurity Support Networks}
These questions asked whether participants had trusted persons, such as family members, friends, or professional networks, to turn to for cybersecurity advice. This section aimed to understand how participants cope with cybersecurity issues and where they seek guidance when needed.

\paragraph{Adapted Questions from Established Surveys}
To enable benchmarking our sample compared to others, we incorporated items from established instruments: 
\begin{itemize}
    \item Objective ability testing questions from the Pew Research Center \cite{smith2017public}. We included 7 questions from the Pew survey ability-based questions provide objective assessment about knowledge on key cybersecurity concepts such as multi-factor authentication, phishing, password security, and encryption. 
    %Each question correctly answered was considered to be worth 1 point yielding a range of 0 to 7 possible scores per participant with higher scores indicating greater objective knowledge on tested cybersecurity concepts. (See Appendix).
        \item The Security behaviour Intention Scale (SeBIS) assesses intentions \cite{egelman2015scaling}, which are related to behaviour \cite{10.1145/2858036.2858265}. It examines secure practices across key domains: password management, account security, and awareness of online threats. Each item was rated on a five-point Likert scale. Responses are summed for an overall SeBIS score.  
        %Participant SeBIS scores range from 20\% to 90\% with higher score signifying high adherence to secure cybersecurity behaviours and lower scores otherwise. (See Appendix).
     \item   An adapted General Self-Efficacy Scale \cite{schwarzer1995generalized} to gauge confidence in handling cybersecurity challenges. The scale comprises ten items rated on a seven-point Likert scale, yielding a composite General Self-Efficacy (GSE).% score ranging from 10 to 70, with higher scores indicating greater levels of cybersecurity self-efficacy (See Appendix). 
\end{itemize}

\paragraph{Demographic Information}
The questions included age, gender, educational background, and other relevant factors that could influence their knowledge and attitudes about cybersecurity.

\subsection{Data Analysis}
\subsubsection{Thematic Analysis} 
Thematic analysis \cite{braun2006using}, an inductive approach widely used in behavioural science due to its flexibility and nuance, was utilised at two levels: at the level of the open-ended questions and at the level of organising all the survey results into themes. To analyse themes in the answers to open-ended questions, we engaged in data familiarization, coding, theme generation, and validation \cite{braun2006using,nowell2017thematic,lincoln1985naturalistic}. The first author immersed himself in the data by repeatedly reading and re-reading the open-ended responses. Weekly research meetings were held where the first author updated the team on the coding progress using a peer debriefing process \cite{lincoln1985naturalistic}. The other authors reviewed and validated the codes and themes during these meetings, reaching consensus through discussion. 

To identify themes across the qualitative and quantitative data, the author team engaged in a similar reflexive process of discussing results and grouping results. We examine the descriptive statistics of the quantitative data holistically in the context of all of the results. The four major themes emerged as we reflected on implications from all the results obtained.  The last theme involved patterns in attitudes and included more targeted statistical analyses which we describe next.

\subsubsection{Analysis of Attitudes}
To determine the strongest underlying concepts in the attitudes, 
we employed exploratory factor analysis. We used the R statistics package and followed the methods of Beaujean \cite{beaujean2013factor}. We first used the ``Scree'' test and determined that three was a suitable number of factors, and then used the ``Varimax'' rotation for exploratory analysis.  We looked at factor loadings, having removed items with magnitudes below $0.5$.
K-means cluster analysis \cite{malik2019applied} was then used to identify distinct subgroups of participants who share patterns in their attitudes.  The ``Elbow'' and ``Silhouette'' \cite{beaujean2013factor} pre-analysis methods suggested two clusters, and K-means with 100 iterations. 2D structuring based on proximity \cite{kassambara2017practical} showed the partitions to be disjoint. We identified two distinct clusters of \textcolor{black}{119 and 117} participants. We then explored how these clusters differed and the implications.

%ADD HYPOTHESES HERE
%We took per-participant means across the questions in each of our subscales, and split the participants according to the two clusters resulting from the K-means analysis to explore how the two clusters were distinct.

\textcolor{black}{Our initial hypotheses explored how the results from our initial questions, SeBIS, Pew, and the adapted GSE, related to the results from our questions about attitude.
In particular, after our factor analysis, we hypothesised that the SeBIS, reflecting intentions, and the Pew results, reflecting knowledge, would correlate positively with feelings of preparedness. Similarly, we hypothesised that the GSE, reflecting self-efficacy, would correlate positively with feelings of confidence. }

\textcolor{black}{After our K-means analysis showed the existence of two clusters in our participant population, we hypothesised that the SeBIS, Pew, and GSE results would show a difference between the two clusters. We also hypothesised that the two clusters would show differences in the sources where participants said they learned about cybersecurity. Finally, based on the studies of students we reviewed in section \ref{sec:relatedwork}, where it was suggested that students had a knowledge-practice gap, we hypothesized that might explain the difference between the two clusters identified.}

\section{Results}
\label{sec:Results}
Thematic analysis identified four overarching themes in computing students' abilities, attitudes and experiences across the qualitative and quantitative data: 
\begin{enumerate}
    \item Computing students are knowledgeable yet have persistent incorrect beliefs.
    \item They learn more from sources outside the classroom.
    \item They have limited assistance and have fallen victim to cybercrime.
    \item Many are confident yet others have mixed concerns about safety and responsibility. Below, we describe the themes, drawing on thematic analysis of the open-ended questions and reporting distributions from the quantitative scales.
\end{enumerate}

\subsection{Computing Students are Knowledgeable, Yet Have Persistent Incorrect Beliefs}

Computing students felt aware of online fraud, with 90\% agreeing or strongly agreeing that they had heard a lot about it. They felt confident in their ability to protect themselves, with 77\% agreeing or strongly agreeing. They reported encountering  deceptive communications: 
\begin{itemize}
    \item 77\% had received deceptive emails,
        \item 88\% had received deceptive SMS text messages,
            \item 47\% had received deceptive phone calls, and
            \item 60\% had received deceptive private or direct messages on social media platforms.
\end{itemize}  

Students' perceptions of their own awareness were substantiated by their ability to correctly point out aspects of digital communications that might warrant suspicion and deception attempts. They were adept at describing the tactics employed by cybercriminals to make their fraudulent schemes appear more convincing and explaining how they deceive individuals into divulging sensitive information. Most emphasised receiving deceptive SMS text messages while only about half (47\%) had received deceptive phone calls, suggesting that text messaging may be a more common vector for scam attempts across our sample. Lastly, 65\% indicated they had been asked to provide information online that they were uncomfortable sharing. Compared to data from National Cybersecurity Alliance, 60\% of their sample had experienced phishing messages (p 16), while 67\% say they can protect themselves (p 18) \cite{OhBehave2025}. For example: 

\begin{quote}\it
One tactic is creating websites that closely resemble legitimate ones, complete with logos and designs to deceive victims. They also use an urgent tone, often threatening severe consequences to instil fear and prompt immediate action. (P 15)

\medskip
Using the same URL as a reputable brand i.e ASB or ANZ but changing the `a' character to a different unicode character thus redirecting the user to a phishing site. (P 45)

\medskip
They make email names to sound just like company names such as paypalsupport@gmail.com, which is obviously not a real email name but sounds convincing. (P 178)
\end{quote}

Participants were well-versed in identifying sensitive data that should be protected and not shared online. They recognised the importance of safeguarding personal details and login credentials, emphasising the need for vigilance online. For example,  participants stated:

\begin{quote}\it 
Naturally, things like username and password obviously are things that I would never give out. I try to avoid purchasing things online, so I suppose I'm cautious about my card credentials.

\medskip
Personal identifiers such as date of birth, home address, or financial information like credit card numbers, bank account information or passwords. Also, sharing my real-time location through GPS can compromise my privacy and safety. (P 211)

\medskip
Passwords and security codes. While I value my private information, I know it's out there due to Instagram, Linkedin and Facebook being repositories used by basically everyone. So it's more important to safeguard the information that I use to access these specific accounts since I do not want a threat actor acting on my behalf at all. I use 2FA on everything and I will never disclose any of this information to anyone online. (P 123)
\end{quote}

\subsubsection{High Accuracy in Objective Knowledge Assessment}

\begin{figure}
\centering
        \includegraphics[width=3in]%,trim={0 6.5cm 0 0},clip]
        {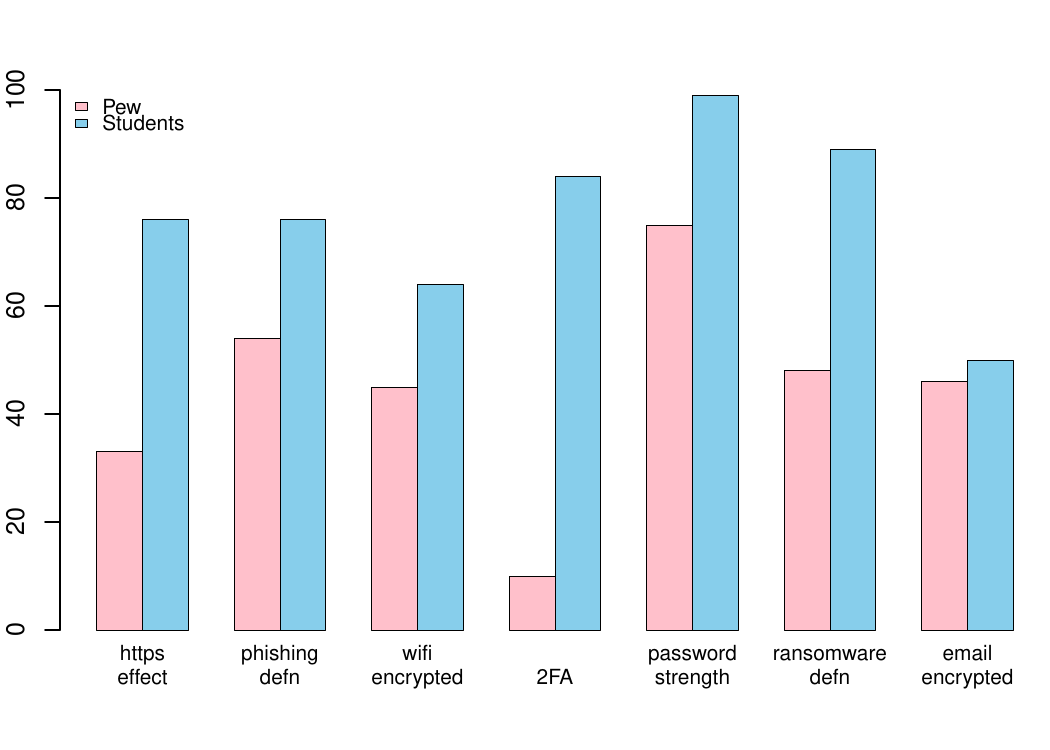}
        \caption{Percentages of participants correctly answering each question, compared with participants in the Pew study. Full text of questions is shown in the appendix.}
        \label{pew}
\end{figure}

%We used seven questions from a Pew survey of more than a thousand participants \cite{smith2017public}.
In Figure \ref{pew} we show the percentage of participants who correctly answered each question, showing data from both the original Pew survey (N = 1000) and our sample. Overall, more students were correct.  Over 75\% of the students correctly answered five or more out of the seven selected questions. The highest concentration of participants scored between 5 and 6 (out of 7 questions) on the scale, while less than 18\% correctly answered only four questions, and 7\% scored less than 4. In comparison, in the original Pew survey \cite{smith2017public}, only one of these questions was correctly answered by 75\% of the participants. (The questions, with correct answers highlighted, are in the appendix.)

%In comparison, in the original Pew survey of more than a thousand participants% \cite{smith2017public}, only one of these questions was correctly answered by 75\% of the participants.

\begin{figure}
\centering

%[trim={left bottom right top},clip]
        \includegraphics[width=3 in,trim={0 6.5cm 0 7cm},clip]
        {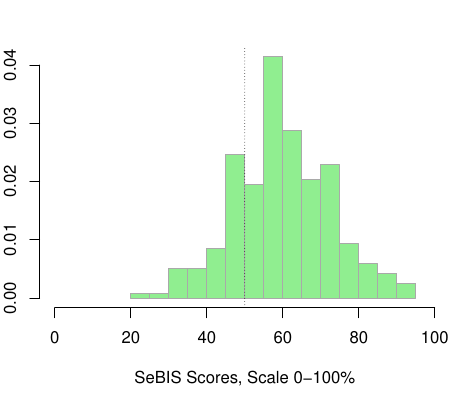}
        
       \caption{Distributions of SeBIS scores, with participants reporting moderate to good adherence to 16 security practices; a score of 100\% would indicate strong agreement with secure behaviours. The dotted lines indicate the midpoint of possible scores. 
       %Each of the questions was answered on a scale from 1 to 5, for each participant, added up to obtain the SeBIS scores as further explained in the appendix.%
       }
        \label{pew-sebis}
\end{figure}

%The range of scores suggests some variability in cybersecurity abilities among the respondents. While many were well-versed across these cybersecurity concepts, others may be at greater risk. %Compared to the result from the PEW survey with a representative cross-section of the American population \cite{smith2017public}, our participants performed overall better. For the seven PEW questions in our study, the median score was 5 out of 7. In the original study, fewer than half the participants scored more than 2 out of 7. 

\subsubsection{Generally Good Cybersecurity Intentions}

%The SeBIS questionnaire examines intentions for secure intentions across several key domains, such as password management, account security, and awareness of online threats. 

Students reported good adherence to secure cybersecurity practices (Figure \ref{pew-sebis}).  While the SeBIS hasn't been used for other student samples, a study with U.S. employees in a range of sectors \cite{umeugo2023security} shows means over 70\% which is higher than the mean for our students: 60\%. However, differing contexts may affect responses: employees use desktop computers for work in a shared office, whereas the students typically use laptops for both coursework and personal use. 

\begin{figure}
\centering
        \includegraphics[width=3.2in,trim={4mm 0 2.5mm 0},clip]{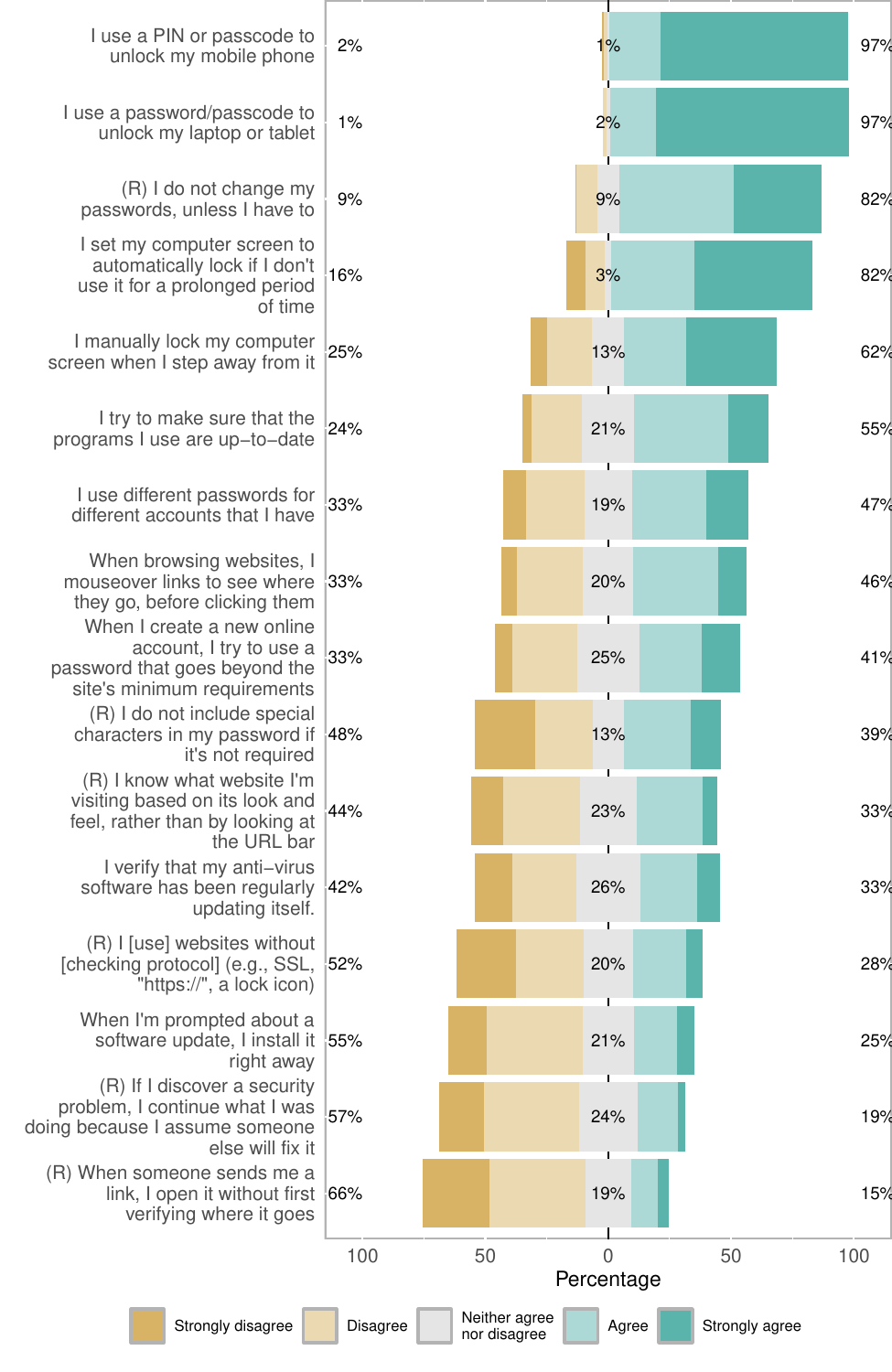}
        \caption{Distributions of responses to questions to the 16 item SeBIS questionnaire. In calculating the overall score, items marked (R) have reversed scores.}
        \label{SeBIS responses}
\end{figure}

The SeBIS indicates overall pro-active cybersecurity behaviour and it also highlights weak practices (Figure \ref{SeBIS responses}). 39\% agreed ``I do not include special characters in my password if it's not required'';  33\% agreed ``I know what website I'm visiting based on its look''.  The ``worst'' response from the students was the 82\% who said they do not change passwords unless they must, but it is not clear how unreasonable that really is. Compared to 2015 when the SeBIS was published, today people have so many passwords that total uniqueness may have become unreasonable, and even experts report some form of password reuse \cite{stobert2018password}.   

%The ``https'' indicator only ever showed connection encryption, not site legitimacy, and since 2019 most phishing websites use free TLS/https certificates \cite{apwg}. Phishing websites are typically short-lived, and even search engine blocklists are can be unreliable \cite{bell2020analysis}. 

\medskip
When asked the practical question about how they would report a suspected online fraud or scam, over half of the participants were unsure what to do. This was particularly true for situations where personal information, such as a home address, phone number, pictures, etc., was compromised rather than login credentials. Noticeable amongst such responses was a sense of helplessness. A few examples of the responses are:

\begin{quote}\it
Depends on the information. If it was banking information, I'd call my bank to lock my cards/accounts. If it was information regarding where I live, I'm not sure. Probably just hope nothing malicious gets done with that information. (P 128)

\medskip
First step I would try to delete the information that I have revealed but knowing what's on the internet is on the internet somewhere, I would not be able to do anything else as its likely beyond my reach. (P 136)

\medskip
I would recall what information I gave out and try to change it if possible, like passwords. If it's my contact or personal details I wouldn't know what to do. (P 196)

\medskip
If it involves a password, I would immediately delete it, change the password to the account, and log out all users. If someone were to store a picture of myself, or my address details, I have no clue how I would control it. (P 68)
\end{quote}

Others mentioned they would contact relevant authorities using platform reporting mechanisms (for example, those offered on social media platforms and email clients), and involve the police.

\subsubsection{Persistent Incorrect Beliefs} 

Certain incorrect beliefs were present. For instance, when asked how they decide whether or not to share information/data, they reported primarily relying on the trustworthiness and reputation of the platform (e.g., Microsoft, Amazon). They looked at: whether a site/message was \textit{``professional-looking''}, security features such as \texttt{https}. Others simply relied on their instincts. Overall: students' trust judgments were based on cues no longer reliable to definitively guarantee trustworthiness. 

%MOVED TO DISCUSSION Today, \texttt{https} does not guarantee the trustworthiness of a website; it only secures the connection. Malicious actors could as well set up secure connection leading to destinations with ill intentions. Secondly, it is easy for cybercriminals to create realistic websites \cite{10.1145/1164394.1164398}, \cite{Ambashtha2023}, replicate existing ones, and generate \textit{``professional-looking''} digital communications, suggesting a critical error stemming from a ``halo'' effect, as also identified in other studies \cite{stojmenovic2022beautiful}. These beliefs can be regarded as obsolete or insufficient in safeguarding against contemporary cybersecurity threats. Adherence to such misconceptions could inadvertently compromise technology users. 

\subsection{Computing Students Learn More From Sources Outside The Classroom}

\begin{figure}
        %\centering
        \includegraphics[width=3.2in,]
        {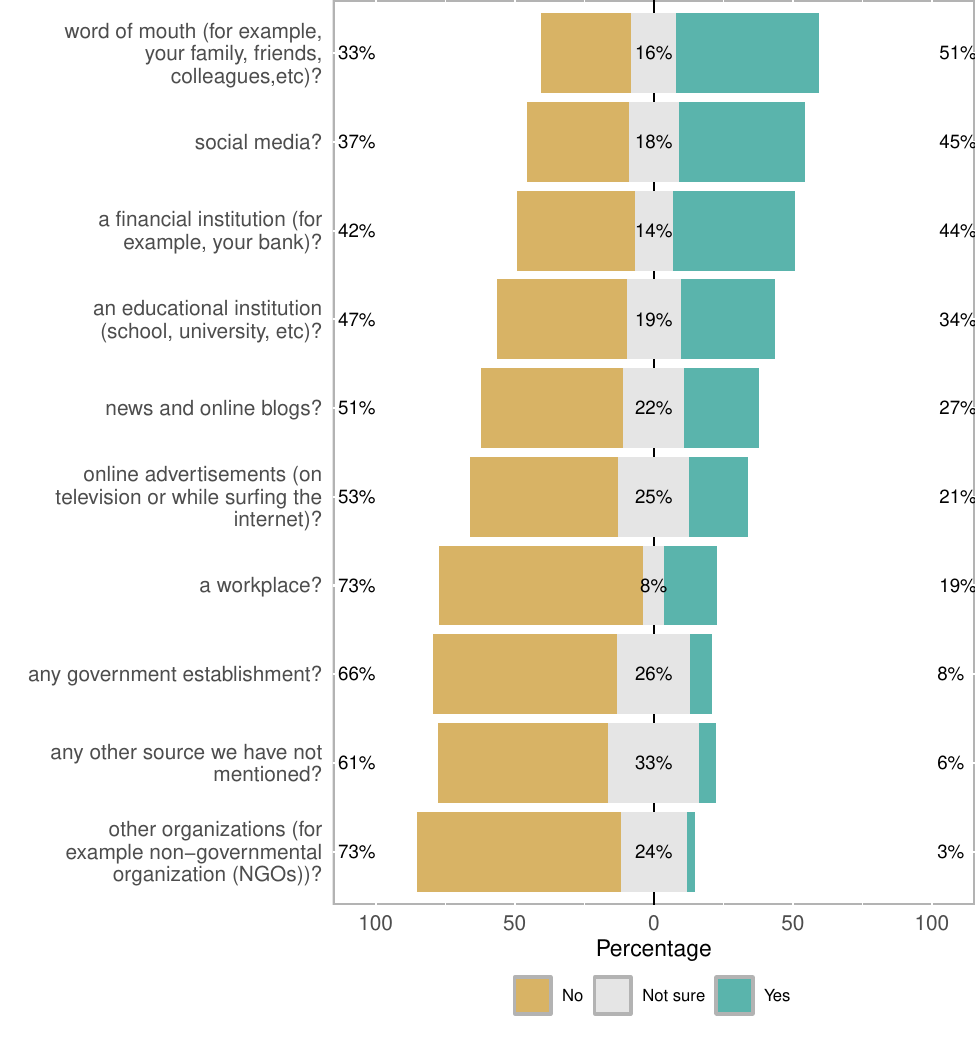}
        \caption{Distributions of responses to possible sources of participants knowledge,  with word of mouth and social media being more prominent sources of knowledge compared to formal institutions.}
        \label{sources}
\end{figure}

Our participants reported learning about scams, warnings, password security, phishing, two-factor authentication (2FA), general cybersecurity practices, and identity protection. The most common sources of knowledge were word of mouth, social media, and financial institutions, with 51\%, 45\%, and 44\% respectively (Figure \ref{sources}).  Of the social media, Reddit was mentioned the most frequently, followed by YouTube, Facebook, TikTok and others. Participants learned about protecting sensitive information from banks and about experiences of victims from word of mouth and social media. Examples of their responses are: 

\begin{quote}\it 
Everything I know about online scams and stuff, my dad taught me.

\medskip
Friends and family have talked about their personal experiences with times they have almost been scammed or any weird texts that they have received. This gave me more knowledge on strategies of scammers and more information on what to look out for. (P 225)

\medskip
Sometimes I'll see some types of scams going around and how to protect yourself against, but I am sceptical of these since the information is coming from social media. (P 60)

\medskip
Occasionally I receive notifications about recent fraudulent activities targeting members of my bank. It keeps me somewhat informed of what to look out for, but it isn't extensive. (P 50)
\end{quote}

Notably, academic institutions were not the primary source of their knowledge. Only 34\% indicated these were a source of knowledge. Few students reported gaining knowledge from news outlets (27\%), online advertisements (21\%) and their workplaces (19\%). Government establishments and NGOs  were also less frequent sources of knowledge, with most not gaining knowledge from these sources or being unsure about it.

% [trim={left bottom right top},clip]

\subsection{Computing Students Have Limited Assistance and Have Experienced Falling Victim to Cybercrime}

Most participants (61\%) did not have trusted persons to whom to turn for cybersecurity help or advice. Only 19\% indicated having someone they trust, and only 13\% reported having a ``Go-To'' person with IT expertise/qualifications. The frequency with which participants sought help from their trusted person was generally low. 15\% reached out frequently, 11\% weekly and most did so less often. When asked if they keep themselves updated about cybersecurity, 61\% admitted that they do not, while only 39\% reported active engagement with cybersecurity information.

%Despite possessing reasonable levels of cybersecurity knowledge,% 
21\% of participants disclosed having  fallen to at least one cyber scam or fraud. 16\% reported they were not sure if they had fallen victim.  40\% of participants reported knowing someone who had fallen victim to a scam or fraud. When asked to share their experiences of falling victim, the participants reported being scammed in various ways. A few reported falling victim as early as 7 and 8 years old. They reported technical attacks, such as phishing with links to fraudulent websites, as well as more general kinds of fraudulent behaviour. One prevalent type of crime involved online marketplaces where they paid for goods or services and did not get anything in return. Compared to \cite{joinson2023development}, 25\% of their sample reported financial related losses.

For example: 

\begin{quote}\it    
I contacted a seller on FB marketplace, transferred money and attempted to pick up the item, but they blocked me. I contacted the bank, but they said they could not return my money because I authorised the transaction. I contacted the police but they never followed up. (P 42)

\medskip
A person pretended to sell an item, we agreed for the item to be shipped once payment was received but the item was never sent after payment was sent. (P 170)

\medskip
I wanted to buy a mattress online. I paid to the person but they then blocked me. (P 41)
\end{quote}

Another category of cybercrime involved clicking on malicious links. Some examples are as follows: 

\begin{quote}\it
I was expecting a package and received a text telling me to make a customs payment. I had just woken up, so I wasn't thinking straight yet and clicked the link to pay. (P 24)

\medskip
I clicked on a dodgy link, got my entire PC taken over including steam account, discord account and more. They got access to my card number via my steam account. I got everything back in the end but it was scary. (P 157)

\medskip
I was sent a message from a friend to scan a QR code to join a server on Discord, and then a bot took control of my account, and sent the messages to numerous other friends. (P 68)
\end{quote}

The last major category of reported cybercrime involved loss of money due to bank accounts or card compromise. For example:

\begin{quote}\it
I woke up and my bank account had been drained of all its money via PayPal. I couldn't link any particular action I did to how my PayPal account was compromised but PayPal were able to reverse the transaction and get my money back. (P 108)

\medskip
They got access to my bank account and they made payments to themselves but luckily I realized early and contacted my bank, and they sorted everything. (P 44)

\medskip
Money was taken out of my bank account without my consent via saved bank details I had in one of my online shopping accounts. (P 208)
\end{quote}

\subsection{\textcolor{black}{Confident But With Concerns}}
%Yet Had Mixed Concerns About Safety, Believing Responsibility Resides With Citizens As Well As Government

%As an important complement to objective abilities and behavioural intentions, we were interested in self-efficacy and attitudes toward cybersecurity. %

\subsubsection{Many Computing Students Were Confident}
\begin{figure}
\centering
%[trim={left bottom right top},clip]
        \includegraphics[width=3 in,trim={0 0 0 13.5cm},clip]
        {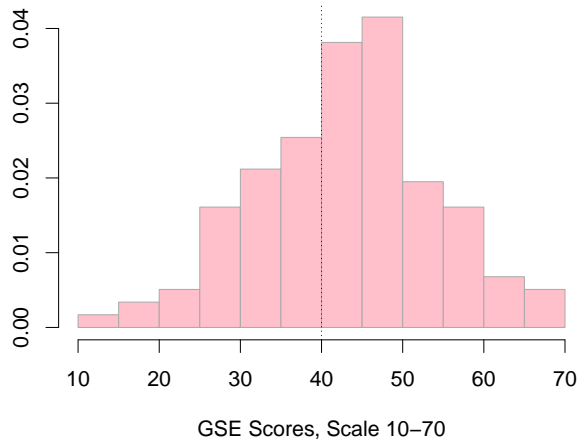}
        \caption{Distribution of scores to the adapted 10 item General Self-Efficacy Scale. The result reflects a roughly normal distribution with a slight skew to the right (positive self-efficacy). The dotted line indicates the midpoint of possible scores.}
        \label{gse}
\end{figure}

\begin{figure}
\centering
        \includegraphics[width=3.5in,trim={5mm 0 0 0},clip]{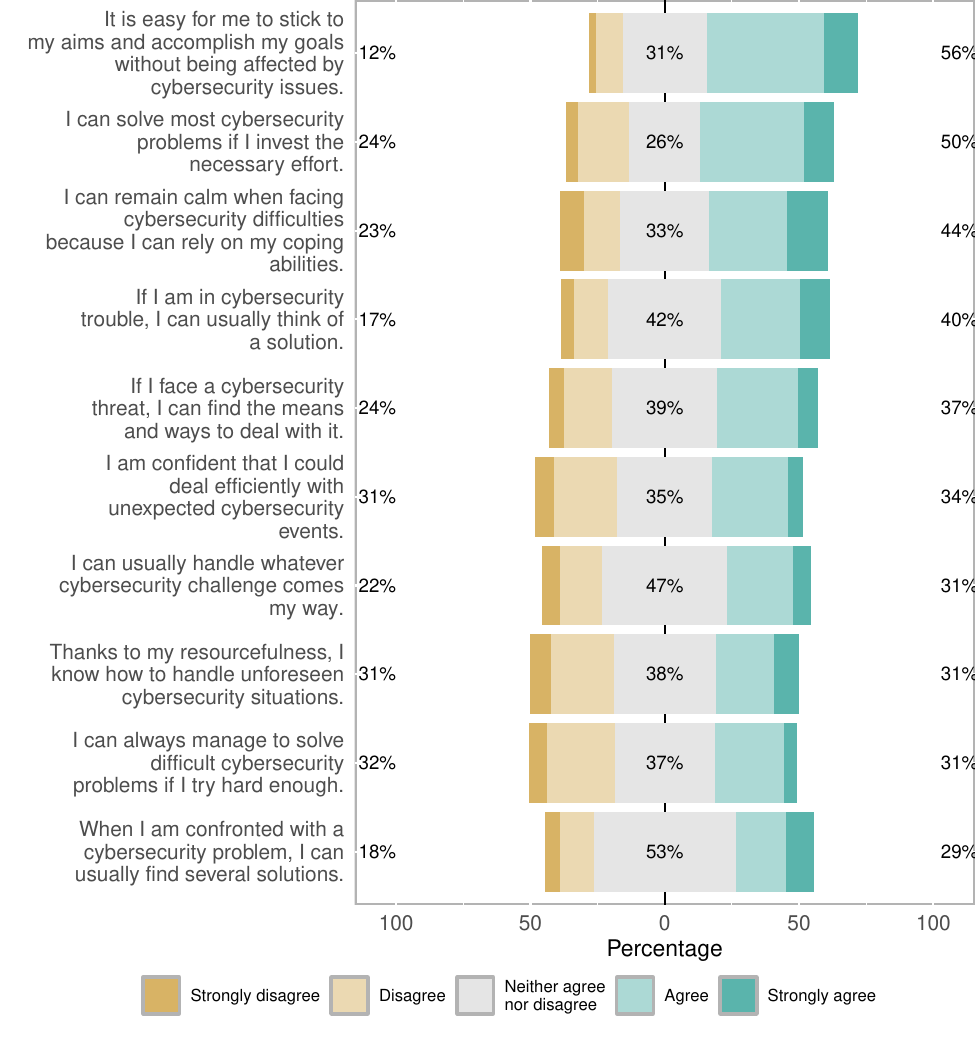}
        \caption{Distribution of responses to questions to the adapted General Self-Efficacy questionnaire.}
        \label{gse_questions}
\end{figure}

Our adapted scale for Self-Efficacy had a roughly normal distribution with a slight skew to the right (Figures \ref{gse} and \ref{gse_questions}). More than half were confident about their self-efficacy and believed they had adequate cybersecurity abilities.  We address confidence further in our analysis of their attitudes, next.

\subsubsection{A Spectrum of Attitudes and Perceptions}

\begin{figure}
%\vspace*{-2.5cm}

\centering
        \includegraphics[trim = 0 0 0 0, clip,width=3.5in]{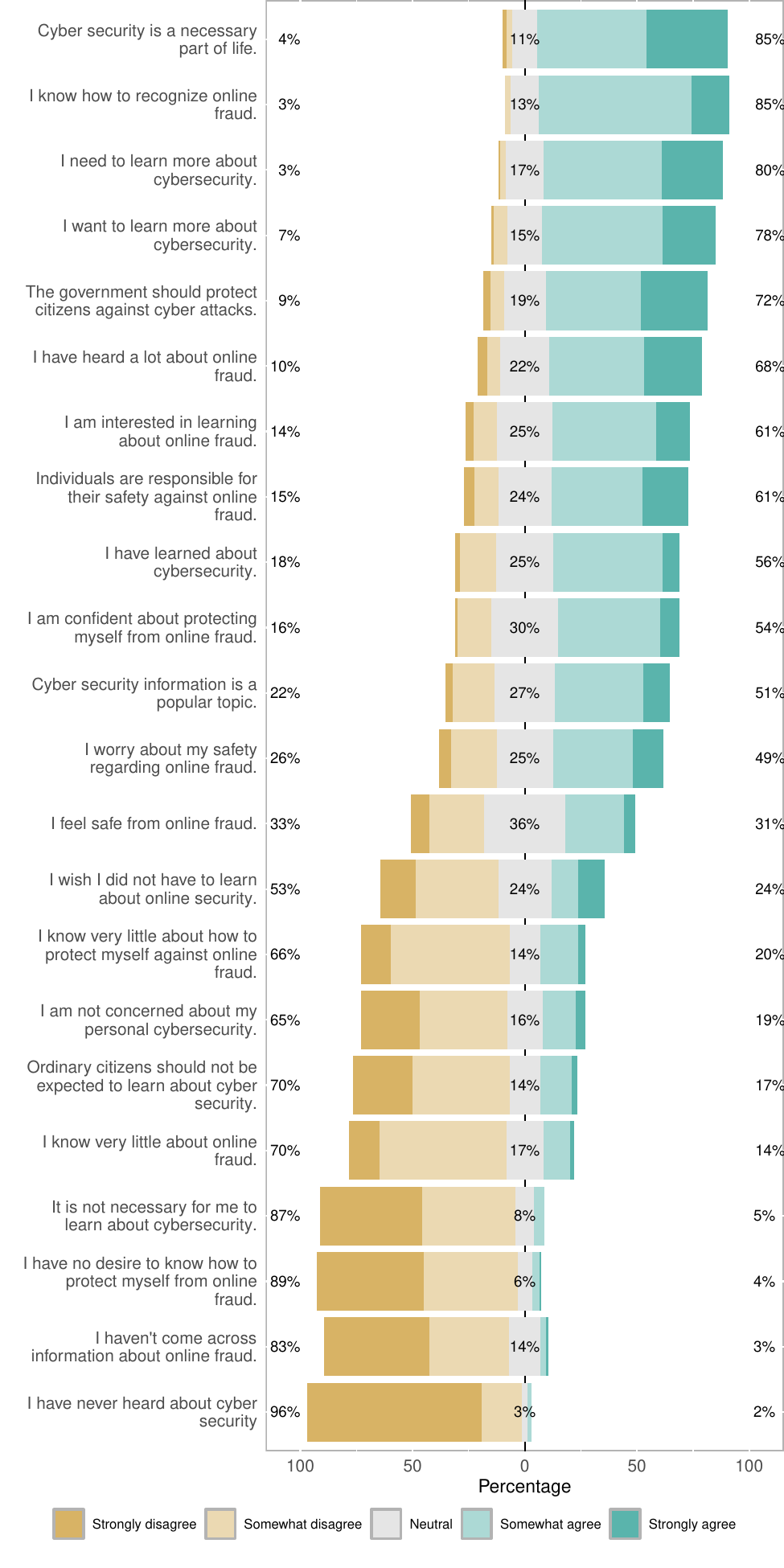}
        \caption{Distributions of question responses to attitudes towards cybersecurity.}
        \label{attitudes}
\end{figure}

Generally, participants reported confidence and engagement, but there are also responses that suggest less engagement (Figure \ref{attitudes}). There was a strong awareness and recognition of the importance of cybersecurity and online protection. 85\% agreed that cybersecurity is a necessary part of life (Figure \ref{attitudes}). About 80\% expressed the need to learn more. There was evidence for beliefs that both individuals and governments should hold responsibility for cybersecurity, with 61\% agreeing that individuals are responsible for their safety and 72\% believing the government should play a protective role.

An exploratory factor analysis identified three factors, which we labelled ``Unprepared'', ``Confident'', and ``Interested'' (Table {\ref{factor-tab}}). They showed acceptable internal consistency with Cronbach's Alphas of .73, .78, and .72, respectively.  The Comparative Fit Index (CFI) was .84. This value is moderate for hypothesised concepts, however it is deemed acceptable for exploratory work. We therefore selected these three factors as a basis for further study.

\medskip

\begin{table}
\caption{Factors from exploratory factor analysis, showing Cronbach's Alpha for each factor, and items with loadings (those with loading magnitudes under $.5$ removed).}
% latex table generated in R 4.5.2 by xtable 1.8-4 package
% Sun Dec 14 07:21:58 2025
\begin{tabular}{|p{6.5cm}|r|}
   \hline
\textbf{Factor: Unprepared $\alpha=0.73$} & Loading \\ 
   I know very little about online fraud. & $0.71$ \\ 
   I know very little about how to protect myself against online fraud. & $0.68$ \\ 
   I have learned about cybersecurity. & $-0.52$ \\ 
   I know how to recognize online fraud. & $-0.56$ \\ 
   \hline
\textbf{Factor: Confident $\alpha=0.78$} & Loading \\ 
   I feel safe from online fraud. & $0.76$ \\ 
   I am confident about protecting myself from online fraud. & $0.63$ \\ 
   I am not concerned about my personal cybersecurity. & $0.61$ \\ 
   I worry about my safety regarding online fraud. & $-0.74$ \\ 
   \hline
\textbf{Factor: Interested $\alpha=0.72$} & Loading \\ 
   I want to learn more about cybersecurity. & $0.81$ \\ 
   I am interested in learning about online fraud. & $0.63$ \\ 
   \hline
\end{tabular}

\label{factor-tab}
\end{table}

\subsubsection{Clusters with Shared Attitudes}

\begin{figure}
%[trim={left bottom right top},clip]
\centering
        \includegraphics[width=\linewidth,trim={0 0 0 0},clip]{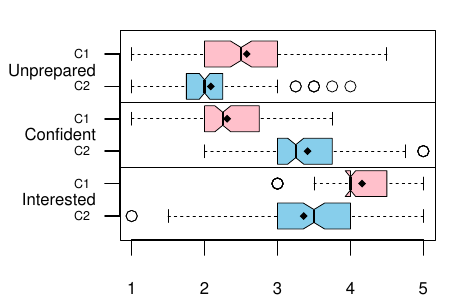}
        \caption[aa]{Two identified clusters across participants, and their factor scores from 1 (Strongly Disagree) to 5 (Strongly Agree). The pink cluster shows a lack of knowledge and capabilities, and less confidence. Interestingly, they are  \textit{more} interested in learning.\\(\small The black diamonds show means, black lines are medians, the coloured boxes represent the inner quartiles, and the whiskers represent the outer quartiles; circles represent outliers. Notches indicate $\approx95\%$ confidence intervals.)}
        \label{clusters-boxplots}
\end{figure}

We applied K-means cluster analysis to explore potential subgroups. Cluster 1 (C1) has higher levels of unpreparedness, low levels of confidence, and higher levels of interest. Cluster 2 (C2) has lower levels of unpreparedness, high levels of confidence, and lower levels of interest in learning more (Figure \ref{clusters-boxplots}). For each factor, the difference between clusters is significant ($p<.0001$).  

In our sample, 119 feel prepared, confident and somewhat interested in learning more and the other 117 feel less prepared, quite unconfident, and very interested in learning more.  

\begin{table*}
\centering
\caption{Comparison of Clusters 1 and 2 by SeBIS, Pew, and GSE scores}
\label{tab:ttests}
% latex table generated in R 4.5.2 by xtable 1.8-8 package
% Tue Mar 10 10:39:46 2026
\begin{tabular}{|l|r|r|r|r|c|}
  \hline
 & $M_1$ & $SD_1$ & $M_2$ & $SD_2$ & Test result with Effect Size \\ 
  \hline
SeBIS & 54.01 & 8.22 & 55.24 & 8.04 & $t(233.99) = -1.16$, $p = .123$, $d = -0.15$ \\ 
  Pew & 5.25 & 1.23 & 5.50 & 1.30 & $W=6064.5, p=0.040$ $r=0.1$ \\ 
  GSE & 39.35 & 10.44 & 47.70 & 10.18 & $t(233.98) = -6.22$, $p < .001$, $d = -0.81$ \\ 
   \hline
\end{tabular}

\end{table*}

\begin{figure}
    \centering
    \includegraphics[width=1\linewidth]{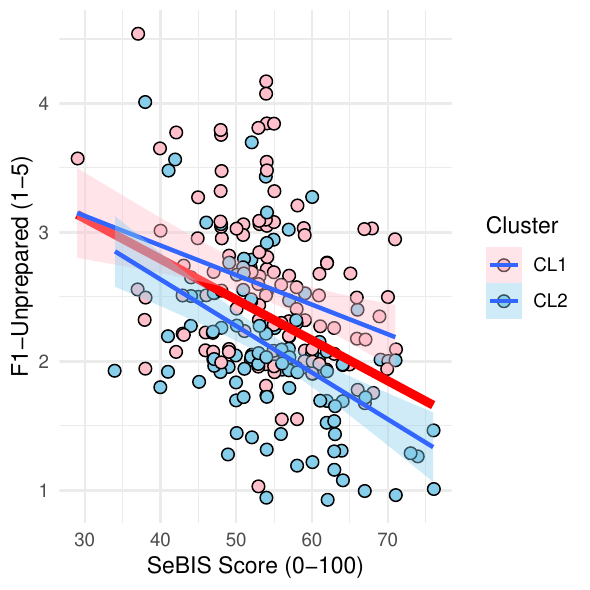}
    \caption{\textcolor{black}{Scatter plot showing relationship between SeBIS score and Factor 1-Unprepared.}\footref{redblue}}
    \label{fig:sebis-f1}
\end{figure}

\begin{figure}
    \centering
    \includegraphics[width=1\linewidth]{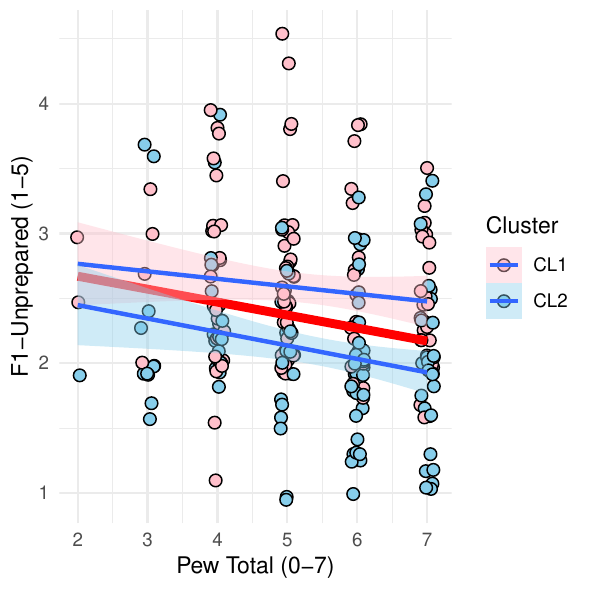}
    \caption{\textcolor{black}{Scatter plot showing relationship between Pew score and Factor 1-Unprepared.} \footref{redblue}}
    \label{fig:pew-f1}
\end{figure}

\begin{figure}
    \centering
    \includegraphics[width=1\linewidth]{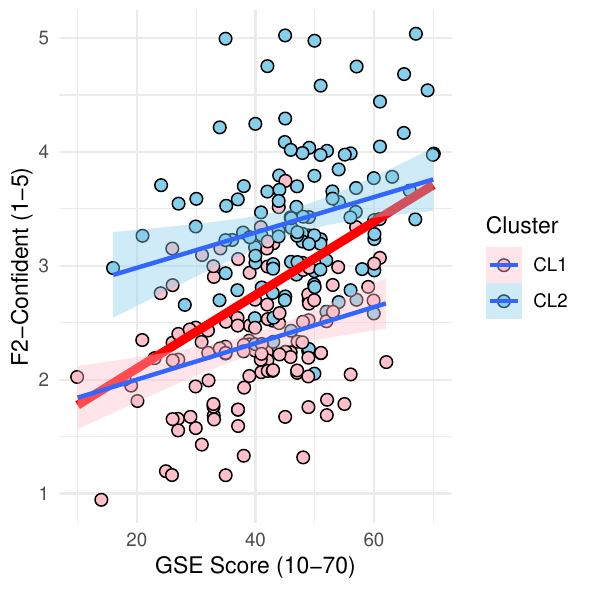}
    \caption{\textcolor{black}{Scatter plot showing relationship between GSE score and Factor 2-Confident. }\footref{redblue}}
    \label{fig:gse-f2}
\end{figure}

\subsubsection{\textcolor{black}{Inferential Analysis}}
\textcolor{black}{
We hypothesised there would be a negative correlation between Security Behaviour Intentions (SeBIS) score and the Pew total, each with Factor 1 (Unpreparedness), and a positive correlation between the adapted General Self-Efficacy (GSE) score and Factor 2 (Confidence). We used Pearson's correlation for SeBIS and GSE, and Spearman's for Pew, because it is a limited ordinal scale. Tests supported these hypotheses: SeBIS-F1: $r(234) = -0.39$, $p<.001$; Pew-F1: $\rho(234) = -0.19$, $p=0.001$; GSE-F2: $r(234) = 0.44$, $p<.001$%
. The tests were conducted for each cluster separately and showed similar significant results. 
Scatter plots for the data are shown in Figures \ref{fig:sebis-f1}, \ref{fig:pew-f1}, and \ref{fig:gse-f2}.\footnote{Red line shows overall correlation, blue lines show correlation for each cluster, along with 95\% confidence bands; points are jittered to better show density.\label{redblue}}}

\textcolor{black}{
We hypothesised that Cluster 1  would score significantly lower than Cluster 2 on SeBIS, PEW, and GSE because the SeBIS would reflect security-conscious behaviour, the Pew would reflect knowledge and a feeling of being well-prepared, and the GSE would reflect feelings of confidence.
We conducted one-sided t-tests for SeBIS and GSE, and a Wilcoxon test for the Pew  (because the distribution was limited and skew). In each case the Cluster 1 means were lower then the Cluster 2 means. The differences between clusters  for Pew and GSE scores showed significant difference, but those for the SeBIS did not: see Table \ref{tab:ttests}. The scatter plot in Figure \ref{fig:sebis-f1} suggests that the SeBIS results for each cluster are mixed, with both low and high SeBIS scores in each cluster.}

\textcolor{black}{
To examine \textit{why} the differences between clusters might have arisen, we considered more specific reasons. For example, whether they had a ``go to'' person, whether that person had expertise,  and whether they had previously fallen victim to cybersecurity attacks, or knew someone who had been.  We found no significant difference between the clusters for any of these.  We then looked at the sources of cybersecurity knowledge reported, hypothesizing differences involving more and less reputable sources. We tested knowledge reported from financial institutions, workplace, education, word-of-mouth, social media, news media, government, and NGOs. Of these, the only significant difference was for word-of-mouth, with Cluster 1 people reporting significantly more use than those in Cluster 2: WoM: $\chi^2(1)=4.641$, $p=0.031$%
 --- without correcting for multiple tests. There also was no evidence that people in each cluster used more sources. Lastly, we explored demographic data and found no significant differences for gender, employment, or age --- almost all students were in the same age range (16--24).}

%Finally, we considered the student degree program. Most of the students were doing Computer Science (CS) degrees, but there were students from a range of other degrees, and many ($\approx 40\%$) students did not identify their degree. We did find a significant difference CS and non-CS students: \input{cs.tex}. While that may offer some insight, we note that there were a similar, but more CS students in cluster 1 than cluster 2, and the significance comes from the non-CS students being much more in cluster 1 than cluster 2: \input{csexpl.tex}. 

\textcolor{black}{To summarise, both intentions (SeBIS) and knowledge (Pew) are related to preparedness (F1) and self-efficacy (GSE) is related to confidence (F2). Yet, intentions do not differ so much as to be significant --- they are much the same. Cluster 1 reported significantly more knowledge coming from Word-of-Mouth.} 

%Perhaps this indicates a need for more engagement with more reputable sources. Are the available sources not targetted to young people, even if studying computer science? Or perhaps are those sources not sufficiently actionable? These questions should inform future research. 

%In the review of related work, Section \ref{sec:relatedwork}, there was the suggestion that although Computer Science students had good knowledge, they had a knowledge-practice gap. This might mean a lack of cybersecurity``know-how'', the kind of  knowledge that indicates ability to achieve success in practical contexts. This reminded us about Bandura's definition of self-efficacy: ``\textit{Perceived self-efficacy is concerned with judgments of how well one can execute courses of action required to deal with prospective situations}'' \cite{bandura1982self}.
%We reflected on this, and reviewed the differences between Cluster 1 and Cluster 2.  
\textcolor{black}{Our final hypothesis concerned the essential difference between the two clusters. We noted that the largest difference between clusters is Factor 2: Confidence. We therefore hypothesised that our GSE score might not only correlate to Factor 2: Confidence, but also explain the distinction between the clusters. We conducted a binary logistic regression with GSE score as the predictor, and cluster as the outcome. We found a significant result ($\chi^2(234)=-3.4962, p<.0001; OR=1.08, 95\% CI=[1.053,1.117]$%
). We also conducted the same test on SeBIS and Pew results, with no significant results. We therefore conclude the results with empirical evidence that \textit{self-efficacy} may play a foundational role in explaining the difference between the two clusters we identified.}

\section{Discussion}
\label{sec:discussion}

\textcolor{black}{Given that today's young people are immersed in technology with complex privacy concerns, are tech savvy yet susceptible and exhibiting gaps between security knowledge and practice, our research question asked: What are computing students’ abilities, attitudes, and experiences around deceptive online interactions, particularly toward phishing? We anticipated that the students' background in computing would show strong abilities. While this was true to some extent, the findings revealed unexpected inconsistencies. The students were more knowledgeable than the general population, but academic courses were not the primary source of their cybersecurity knowledge. We found they lacked adequate support systems, and some had fallen victim to cybercrime. They were mostly confident about their abilities, but they also expressed notable concerns about their safety. }

\subsection{\textcolor{black}{Contributions and Association to Theory}}

% In section \ref{sec:complacency}, we address their complacency, and theories such as digital resignation, fatigue, and cynicism. Finally, in section \ref{sec:education}, we consider the nature of cybersecurity education and theories such as social learning and self-efficacy.

\textcolor{black}{In our cluster analysis, one set of participants felt less prepared and less confident, while the other felt more prepared and more confident. A well-known but surprising connection between competence and confidence is the ``Dunning-Kruger effect'', where ``poor performers in many social and intellectual domains seem largely unaware of just how deficient their expertise is''\cite{DUNNING2011247}.  We studied relationships between objective knowledge and confidence in our study, but did not find any evidence that suggested people with less knowledge were more confident. However, it is interesting that our participants with less confidence were on average more keen to learn.}

\textcolor{black}{A relevant theory of attitudes is ``Security Fatigue'' where one feels constantly vigilant online, bombarded with cybersecurity warnings, password requirements, and threat alerts, contributing to fatigue and complacency \cite{furnell2009recognising}. This attitude was observed in a workplace context as well as in everyday life \cite{stanton2016security}, for example: 
\begin{quote}
I think I am desensitized to it—I know bad things can happen. You get this warning that some virus is going to attack your computer, and you get a bunch of emails that say don’t open any emails, blah, blah, blah. I think I don’t pay any attention to those things anymore because it’s in the past. People get weary of being bombarded by ``watch out for this or watch out for that'' (participant 101 (p. 26)\cite{stanton2016security}) 
\end{quote}}
\textcolor{black}{A related theory ``compliance budget'' also originates in the workplace and describes how employees weigh the costs and benefits of compliance. When their limit is reached, individuals may refuse to comply or find ways to circumvent requirements, as their willingness declines with the introduction of additional security policies and demands \cite{beautement2008compliance}.  Students are not under the same demands as employees, yet we observed correlations of behaviour intentions and objective security knowledge to preparedness and confidence. Our logistic regressions points to the deeper source of self-efficacy: ``\textit{Perceived self-efficacy is concerned with judgments of how well one can execute courses of action required to deal with prospective situations}'' \cite{bandura1982self}.}

\textcolor{black}{The phenomena around “digital resignation” \cite{draper2019corporate} and ``privacy cynicism'' carry some similarities to what we observed in our wide cross section of data on attitudes towards cyber security. For privacy, there is a sense of young people feel resigned to having their data tracked and they feel like any efforts would be ``futile.''  This suggests self-efficacy could be a moot point for some, as asking about self-efficacy for something impossible fundamentally does not make sense. Because we observed interest in cybersecurity alongside a lack of confidence and self-efficacy, this suggests there could possibly also be an emerging attitude of resignation for cyber security; our research suggests this would be important future work.}

\textcolor{black}{We observed that those in Cluster 1 (unprepared, lacking confidence, interest in learning) were more likely to learn through word-of-mouth, which is relevant to the ``social learning'' theory \cite{das2014effect}. The reliance on informal sources could relate to the broader issues of confidence --- students may feel confusion or a lack of clarity in information coming from these sources, which ultimately may not be increasing their security self-efficacy. We discuss more practical considerations regarding confidence next.}

\subsection{Confidence: Why Do Students Feel Confident Despite Having Cybersecurity Concerns}

It is odd that confidence in the cybersecurity abilities of one of our clusters coexists with significant worries in the same cohort. One possible explanation is the fact that cybercrime is constantly taking different forms and increasing in complexity \cite{murphy2024understanding}. Even otherwise confident students might recognise that their abilities have limitations, especially in the face of increasingly sophisticated threats. Automaticity may play a role, where users may behave ``automatically'', like clicking on phishing links \cite{Vishwanath2018suspicion}. Especially if they become victims this way, they may feel that they cannot even trust themselves. 

Students may feel prepared to handle known risks, but are aware that new and unforeseen dangers can emerge (for example, ``zero day'' vulnerability attacks \cite{bilge2012before}). This awareness can create a sense of vulnerability, where confidence in current abilities does not fully alleviate concerns about unknown threats. In comments, computing students rightly noted the significant impact that artificial intelligence could have on cybercrime. 

\begin{quote}\it
    AI also is most likely going to prove itself to be able to improve scammers ability to conduct fraud to, and possible to a point where even knowledgeable individuals could fall for it. (P 159)
    
\medskip

    Fake information would appear more frequently, and due to the increasing usage of AI recently, people can create real voices and real images and use them for online fraud in new ways. (P 162)
\medskip

That scammers are always developing their tactics, and in the future could involve AI that can accurately impersonate relatives/friends. (P 8)
\end{quote}

With generative AI able to create legitimate looking communications with both customization, variation, and perhaps even interaction, such threats should be taken seriously \cite{heiding2024devising,weinz2025impact,roy2024chatbots}.

Secure practices seemingly require users to scrutinize online communications with the attentiveness of an expert without clear or consistent guidance \cite{cyberdb2025bestpractices,tenfour2025securecommunication}. Common cues of legitimacy can a times be unreliable, security advice is often impractical or conflicting, and users are expected to manage tasks that exceed human capabilities. As these challenges accumulate, the burden of staying secure grows, perhaps raising concerns amongst computing students, as shown in previous quotes. Additionally, secure practices may become obsolete. We found that students' trust judgments for the legitimacy were based on cues no longer reliable to definitively guarantee the trustworthiness of certain digital communications.

% NEEDS TO BE MEVED SOME WHERE ELSE Similarly, as pointed out by our participants in their open ended responses, it is easy for cybercriminals to create realistic websites \cite{10.1145/1164394.1164398}, \cite{Ambashtha2023}, replicate existing ones , and generate \textit{``professional-looking''} digital communications. Today, relying on how professional a digital communication looks could lead to a critical error stemming from a ``halo'' effect, as also identified in other studies \cite{stojmenovic2022beautiful}.  The same applies to communications without grammatical flaws which today could easily be generated by large language models. These heuristic trust judgments can today be regarded as obsolete or insufficient in safeguarding against contemporary cybersecurity threats. Adherence to such misconceptions could inadvertently compromise technology users. 

\subsection{Complacency: Why Do Some Express Lack of Interest In Learning About Cybersecurity}
\label{sec:complacency}
A substantial portion of students were not particularly interested in learning more about cybersecurity and improving their abilities; they were more confident yet near-neutral in interest. This is particularly surprising as we anticipated a more engaged posture owing to their IT inclination.

A plausible reason could be a lack of time or resources involved in developing a good cybersecurity posture. Many individuals, such as students, might already be balancing demanding academic schedules, work, and personal commitments, leaving little room for additional learning activities. Cybersecurity might not be considered an immediate priority in this context unless they have faced personal cyber incidents. Results from our factor analysis (``I want to learn more'' factor 3) suggests that while some acknowledge the importance of cybersecurity, they struggle to find the capacity to invest the time and resources necessary to develop their abilities\textcolor{black}{, which would be relevant to the theory of ``compliance budget''. } Unlike other well-established aspects of our lives where we outsource certain responsibilities, such as handing our cars to a mechanic to fix, or calling on a plumber to fix a leaking pipe, personal cybersecurity cannot be delegated.  Moreover, there may be an effect of attitudes like ``digital resignation'', or ``cynicism'', as observed for privacy, on interest in learning about security: some students may simply be accepting that they cannot be fully secure, which unclear implications for learning. For example, students in our sample stated:

\begin{quote}\it

I don't know if my efforts to be vigilant are enough or will be enough in the future. I am aware of hacking and that sometimes things are just not in our control. (P 54)

\medskip

I can't keep up to date with ongoing scams all the time, and there may be moments where I am not in the headspace to think critically. (P 3)

\medskip

Kind of annoying to keep on tracking new ways that people find ways to hack and online fraud and you have to keep updating your security and learn ways that people try to do fraud on you. (P 77)

\medskip

How careful do I need to be? Is there a certain number of steps I need to take to ensure I am fully protected? (P 33)

\medskip

Why is it something that someone needs to do, rather than have a standard that prevents it from happening so often (P 31)
    
\end{quote}

\textcolor{black}{At the outset of this study, we were not anticipating finding attitudes related to cyber security anxiety or resignation. Emerging work coming out this year is showing cyber security anxiety \cite{dall2026fear} which they distinguish as current concerns, future anticipated threats, and perceived control over outcomes. The notion of perceived control relates back to the observed underlying strength of self-efficacy to explain clusters of attitudes and perceptions related to preparedness, confidence, and interest in learning.}

\subsection{What Should Constitute Cybersecurity Education For Students and The Public?}
\label{sec:education}

Our findings show that computing students were knowledgeable, yet, gaining this knowledge from sources other than academic courses raises several questions. Current standards for cybersecurity tertiary education are to prepare graduates to ``design and develop more secure code, ensure data security and privacy, and apply a security mindset to their daily activities'' (p. 256, \cite{servin2024cs2023}) yet there is a disconnect between their formal education and the practical knowledge they possess; this was also observed by others \cite{moallem2019cybersecurity}. One potential reason could be the difficulty for academic institutions to keep course content up-to-date with new cybersecurity threats and vulnerabilities. Or, to remain consistent, academic courses might be overly theoretical, emphasising fundamental concepts and abstract theories, with little emphasis on practical, hands-on skills that of importance.  As a result, students may not be fully prepared for the dynamic and unpredictable nature of cyber-threats in the real world.

 While students may learn theoretical knowledge about computing and cybersecurity in academic settings, they may not be exposed to more practical knowledge about defending oneself in those settings. They may hear practical stories and lessons from friends and social media, which aligns with work on ``social learning'' \cite{das2014effect}. The reliance on informal sources leading to consuming huge amounts of media coupled with the need for constant vigilance amongst many could consequently point to the broader issues of confidence and motivation.  Students may feel overwhelmed by the sheer volume, diversity, and complexity of cyber-security information online. This can lead to confusion for some and a sense of helplessness for others, not knowing where to start or what information to trust. Sources of cybersecurity information might convey false or contradictory information. As a result, users can selectively engage only with the most clear and accessible sources. This raises questions about how to present cybersecurity information in a clear, reliable, and comprehensive way. Similar observations were made regarding information on securing home IoT devices \cite{turner2021googling}: their findings revealed it is extremely challenging to find coherent, actionable and reputable cybersecurity guidance online. 

We speculate that it is possible that studying computing synergises with  informal sources to improve abilities. These students draw on the same online resources, personal experiences, and real-world encounters as individuals without an IT background, and their academic exposure to computing concepts likely amplifies their ability to better interpret and apply this knowledge. This is in line with research on how computing education makes a difference in cybersecurity awareness between students undertaking a computing-related degree and others \cite{venter2019cyber}.

\textcolor{black}{A promising direction would be a focus on a practical, hands-on curricula that aim to improve users' cybersecurity competencies and self-efficacy. Cybersecurity self-efficacy functions as both an antecedent and outcome of numerous psychological and behavioural variables \cite{borgert2024self}. Interactive and collaborative learning environments can strengthen self-efficacy, thereby encouraging engagement with cybersecurity concepts and practices \cite{sun2022effects}. Educational interventions grounded in self-efficacy principles can significantly improve users’ confidence and attitudes toward using security tools and adopting safe practices \cite{chen2020hacked}.   Collectively, these studies demonstrate that cultivating self-efficacy is fundamental for improving everyday users’ cybersecurity abilities, suggesting that effective cybersecurity education and interventions should prioritise strategies that build users’ confidence, skills, and perceived competence in managing digital threats \cite{stavrou2024cultivating}.}

While improving education and communications about cybersecurity remain important goals, cybersecurity challenges are systemic. Problems stem from the very low cost of deploying cyber attacks and the lack of reliable identification requirements. Organisational responsibilities may have a role to play, rather than leaving individuals to protect themselves alone. 

In summary, our study revealed that even computing students face notable challenges with their personal cybersecurity. They are not immune to mistakes, as shown by inconsistencies in understanding and practice. Furthermore, this reveals the need and importance of adequate support systems and resources, even for such a category. Surprisingly, some students displayed concerning levels of indifference towards cybersecurity. These findings suggest the need to develop better support systems, guidance, and improved educational tools, as well as for more research efforts on the factors that contribute to these attitudes and behaviours.

\subsection{Limitations} 
Self-reported accounts of behaviour are subject to social desirability bias. This is an overarching concern pertaining to security behaviour research. Responses to instruments such as the SeBIS, GSE and certain open-ended questions may not accurately reflect participants' actual behaviours and attitudes, as they can be influenced by social desirability, overconfidence, or limited self-awareness. \textcolor{black}{In our questions on ``attitudes'' we asked a variety of questions about perceptions and self assessment. Future work should more reflect previous research on cybersecurity attitudes specifically such as that by Hadlington \cite{hadlington2017human} and Faklaris \cite{faklaris2019self}.} Also in future studies we aimed to address this concern by conducting follow-up interviews with selected participants to further probe certain aspects. 

Another key limitation of this study lies in its sampling approach, as all participants were drawn from a single institution. While this provided consistency in curriculum exposure and other contexts, it may limit the generalizability of the findings to broader populations of computing students across different universities or regions. Institutional culture, teaching practices, and resource availability may uniquely shape students' experiences and awareness, potentially leading to results that do not reflect the diversity of computing education environments. Future studies could aim to include participants from multiple institutions to enhance the external validity and applicability of the findings. \textcolor{black}{In a study such as this, we also recognise the importance of debriefing and providing an opportunity for students to learn more about the cybersecurity issues. We intended to do this in class, but the survey was not completed until the end of the semester. }

Lastly, personal cybersecurity is quite broad and complex. More studies beyond ours, focusing on different aspects of personal cybersecurity,  leveraging other research approach such -- as additional follow-up interviews, longitudinal studies is critical to corroborate the data from our survey. 

Further studies could explore methods that facilitate the reliable acquisition of cybersecurity abilities, and keep users' cybersecurity awareness current and adaptable. Finally, addressing the lack of adequate support structures is essential. Future research efforts could consider developing and evaluating support systems that facilitate the safe use of technology and assist users in navigating cybersecurity complexities.

\section{Conclusion}
\label{sec:conclusion}

This study involved a comprehensive survey on computing students' cybersecurity attitudes, behaviours, and experiences, which uncovered inconsistencies: 1) they are objectively knowledgeable yet hold out-of-date heuristics for identifying online deception, 2) they learn from non-academic sources, 3) they have a limited support network and have fallen victim to cybercrime, and 4) we discovered a range of preparedness, with one group more enthusiastic, confident yet feeling less safe and another group feeling unsure, unexcited about learning more, and feeling less concerned about their own safety. 

\textcolor{black}{Beyond that, the data and analysis reveal traces of sentiments that resonate with ``digital resignation'' \cite{draper2019corporate}, ``security fatigue'' \cite{furnell2009recognising}, the ``compliance budget'' \cite{beautement2008compliance}. These theories warned against stretching user limits with respect to security. They posit that these sentiments are rational reactions to our current cybersecurity landscape.  }

These findings underscore the need for a two-pronged approach to improving cybersecurity abilities, as some people will be eager and open to learning and acknowledge their need to learn, while others may not be as open nor as concerned. Further research on attitudes, behaviours, and experiences is needed to better understand disparities in cybersecurity interest, knowledge, and practices.

\hfill

\textit{Acknowledgement} \\We want to acknowledge and appreciate the University of Auckland for the support with the gift cards presented to the raffle draw winner.\\

%%% ----- PAPER ENDS HERE

%\begin{equation}
%\end{equation}

% - For one-column wide figures use
% \begin{figure}
 %- Use the relevant command to insert your figure file.
 %- For example, with the graphicx package use
%  \includegraphics{example.eps}
 %- figure caption is below the figure
% \caption{Please write your figure caption here}
% \label{fig:1}       % Give a unique label
% \end{figure}

%
% - For two-column wide figures use
%\begin{figure*}
% - Use the relevant command to insert your figure file.
% - For example, with the graphicx package use
%  \includegraphics[width=0.75\textwidth]{example.eps}
% - figure caption is below the figure
%\end{figure*}

%
% - For tables use
% \begin{table}
% - table caption is above the table
% \caption{Please write your table caption here}
% - For LaTeX tables use
%\begin{tabular}{lll}
%\hline\noalign{\smallskip}
%\end{tabular}
%\end{table}

%\begin{acknowledgements}
%\end{acknowledgements}

\bibliographystyle{spmpscinat}      % mathematics and physical sciences
\bibliography{Bibliography.bib}   % name your BibTeX data base

% - Non-BibTeX users please use
%\begin{thebibliography}{}
%
% - and use \bibitem to create references. Consult the Instructions
% - for authors for reference list style.

% - Format for Journal Reference
% - Format for books
% - etc
%\end{thebibliography}

\includepdf[pages=1-last,frame,offset=0 -0.5cm,noautoscale=false,scale=.8,pagecommand={\thispagestyle{plain}}]{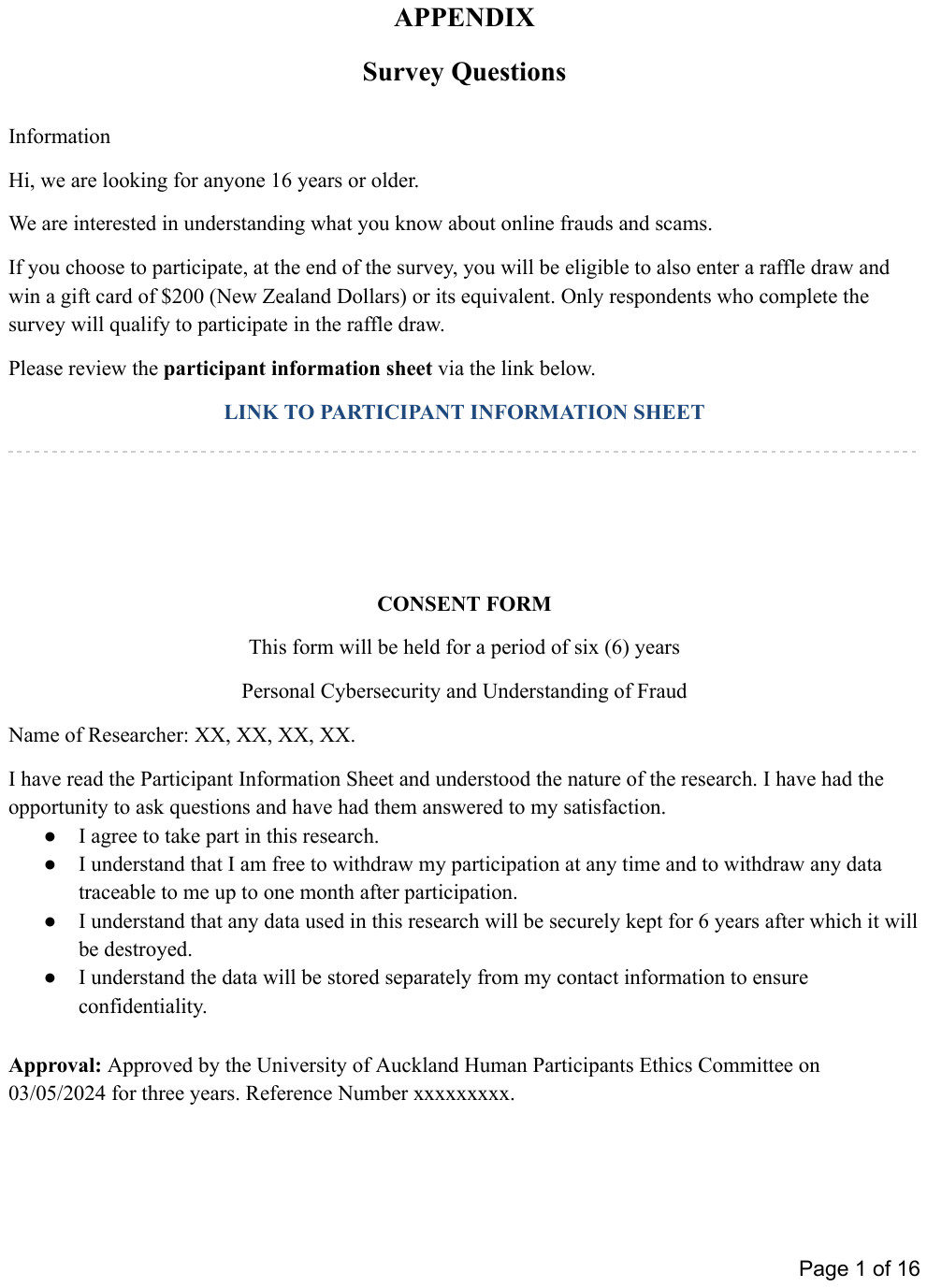}

\end{document}